\documentclass[]{spie}  

\usepackage{amsmath,amsfonts,amssymb}
\usepackage{graphicx}
\usepackage[colorlinks=true, allcolors=blue]{hyperref}

\title{The Roman Coronagraph Community Participation Program: data reduction pipeline design and implementation}

\author[a,b]{Jason J. Wang}
\author[c]{Maxwell A. Millar-Blanchaer}
\author[d]{Marie Ygouf}
\author[d]{Julia Milton}
\author[e]{Jürgen Schreiber}
\author[f]{Kevin J. Ludwick}
\author[g]{Ellis Bogat}
\author[a,b]{Amanda Chavez}
\author[c]{Eric Shen}
\author[a,b]{Aneesh Baburaj}
\author[h]{Ramya Anche}
\author[r,i]{Toshiyuki Mizuki}
\author[r,i]{Taichi Uyama}
\author[k]{Ezar Shinbaro}
\author[m]{Alexis Lau}
\author[l]{Neil T. Zimmerman}
\author[m]{Sophie Noiret}
\author[n]{William Balmer}
\author[o]{Ben J. Sutlieff}
\author[d]{Adrien Maillard}
\author[e]{Matthias Samland}
\author[d]{A J Eldorado Riggs}
\author[p]{Clarissa Do {\'O}}
\author[c]{Jingwen Zhang}
\author[o]{Giovanni M. Strampelli}
\author[m]{Lisa Altinier}
\author[d]{Michael Liang}
\author[q]{Jessica Gersh-Range}
\author[d]{Kylee Fluckiger}
\author[r,i]{John Livingston}
\author[a,b]{Malachi Noel}
\author[s]{Dmitry Savransky}
\author[t]{Michele Woodland}
\author[u]{Tsutsumi Nagai}
\author[h]{Justin Hom}
\author[d]{Al Niessner}
\author[d]{Cynthia Wong}
\author[d]{Eric J. Cady}
\author[d]{Vanessa P. Bailey}
\author[v]{Guillermo Gonzalez}
\author[w]{Alexandra Z. Greenbaum}
\author[g,l,x]{Enrico Biancalani}

\affil[a]{Northwestern University, 633 Clark Street Evanston, IL 60208, USA}
\affil[b]{Center for Interdisciplinary Exploration and Research in Astrophysics, 1800 Sherman Ave, Evanston, IL 60201, USA}
\affil[c]{University of California, Santa Barbara, CA 93106, USA}
\affil[d]{Jet Propulsion Laboratory, California Institute of Technology, 4800 Oak Grove Drive, Pasadena, CA 91109, USA}
\affil[e]{Max Planck Institute for Astronomy, Königstuhl 17, 69117 Heidelberg, Germany}
\affil[f]{University of Alabama in Huntsville, 301 Sparkman Dr, Huntsville, AL 35899, USA}
\affil[g]{University of Maryland College Park, Center for Research and Exploration in Space Science and Technology II (CRESST II), NASA GSFC, 8800 Greenbelt Rd, Greenbelt, MD, USA}
\affil[h]{Steward Observatory and the Department of Astronomy, The University of Arizona, 933 N Cherry Ave, Tucson, AZ, 85721, USA}
\affil[i]{National Astronomical Observatory of Japan, NINS, 2-21-1 Osawa, Mitaka, Tokyo 181-8588, Japan}
\affil[k]{Lake Forest College, 555 North Sheridan Road, Lake Forest, IL 60045, USA}
\affil[l]{NASA’s Goddard Space Flight Center, 8800 Greenbelt Rd., Greenbelt, MD 20771}
\affil[m]{Aix Marseille Univ, CNRS, CNES, LAM, Marseille, France}
\affil[n]{Space Telescope Science Institute, 3700 San Martin Drive, Baltimore, MD 21218, USA}
\affil[o]{Institute for Astronomy, University of Edinburgh, Royal Observatory, Blackford Hill, Edinburgh, EH9 3HJ, UK}
\affil[p]{Caltech, 1200 E. California Blvd., Pasadena, CA, 91125 USA}
\affil[q]{DM Telescopes LLC, Raleigh, NC 27615, USA}
\affil[r]{Astrobiology Center, NINS, 2-21-1 Osawa, Mitaka, Tokyo 181-8588, Japan}
\affil[s]{Sibley School of Mechanical and Aerospace Engineering, Cornell University, Ithaca, NY, 14853, USA}
\affil[t]{Department of Astronomy and Astrophysics, University of California Santa Cruz, 1156 High Street, Santa Cruz, CA 95064, USA}
\affil[u]{The Graduate Univ. for Advanced Studies, Shonan Village, Hayama, Kanagawa 240-0193, Japan}
\affil[v]{Tellus1 Scientific LLC, Huntsville, AL, USA}
\affil[w]{IPAC, Caltech, 1200 E. California Blvd., Pasadena, CA 91125, USA}
\affil[x]{Center for Research and Exploration in Space Science and Technology II (CRESST II), Greenbelt, MD, USA}

\authorinfo{Further author information: (Send correspondence to J.J.W.: jason.wang@northwestern.edu)}

\begin{document} 
\maketitle

\begin{abstract}
The Roman Space Telescope Coronagraph Instrument will demonstrate a series of technologies and techniques to enable the direct detection of reflected-light planets with space-based observatories. To characterize and validate the performance of the Coronagraph Instrument, the Community Participation Program is developing corgidrp, an open-source Python-based data reduction pipeline. The pipeline can process data from the required and best-effort observing modes and their associated calibration sequences into calibrated science-ready data products. We present the software design and implementation of corgidrp and the motivation behind specific design decisions. We describe the software architecture, data flow, processing steps, automation tools, testing framework, and development philosophy. We also outline future development plans in preparation for on-sky data.
\end{abstract}


\section{INTRODUCTION}
\label{sec:intro}  
The Coronagraph Instrument onboard the Roman Space Telescope (henceforth referred to as the ``Roman Coronagraph") aims to demonstrate a series of hardware and observing techniques to unlock new regimes of exoplanet imaging, with the goal of reaching contrasts of $10^{-8}$ or better at visible light wavelengths for the first time \cite{Bailey2023}. The Roman Coronagraph has the potential to image planets and exozodiacal dust in reflected light for the first time. The lessons learned from the Roman Coronagraph will help pave the way for future space missions like the Habitable Worlds Observatory to image Earth-sized planets in the habitable zone of other stars. 

The Roman Coronagraph Community Participation Program (CPP) \cite{Savransky2024} has been assembled to prepare for and execute technology demonstration observations for the Roman Coronagraph. The CPP is organized into working groups to cover different aspects of the effort (e.g., Ref.~\citenum{Wolff2024}). The Data Reduction and Simulations working group leads the data effort, both in modeling what real data will look like and how to process real data \cite{Millar-Blanchaer2024}. The Data Reduction and Simulations working group leads the development of two key open-source packages: \texttt{corgisim} to simulate data products \cite{Zhang2026} and \texttt{corgidrp} to reduce raw data. The \texttt{corgidrp} package is the data reduction pipeline for the Roman Coronagraph. It will produce calibration products needed for the operation of the Roman Coronagraph's wavefront control algorithms. It also will be used to reduce all science data taken with the Roman Coronagraph. Along with the raw data, a default processing of all science data will be publicly released with no proprietary period. An overview of \texttt{corgidrp} can be found in Ref.~\citenum{Millar-Blanchaer2024}. \texttt{corgidrp} is publicly available on Github\footnote{\url{https://github.com/roman-corgi/corgidrp}} and the latest release v5.0 is archived on Zenodo\cite{corgidrpv5}. This proceeding will describe the design and implementation details of the package.

\section{Roman Coronagraph Data Reduction Needs}
\label{sec:needs}
\begin{figure}
    \centering
    \includegraphics[page=5, width=\textwidth]{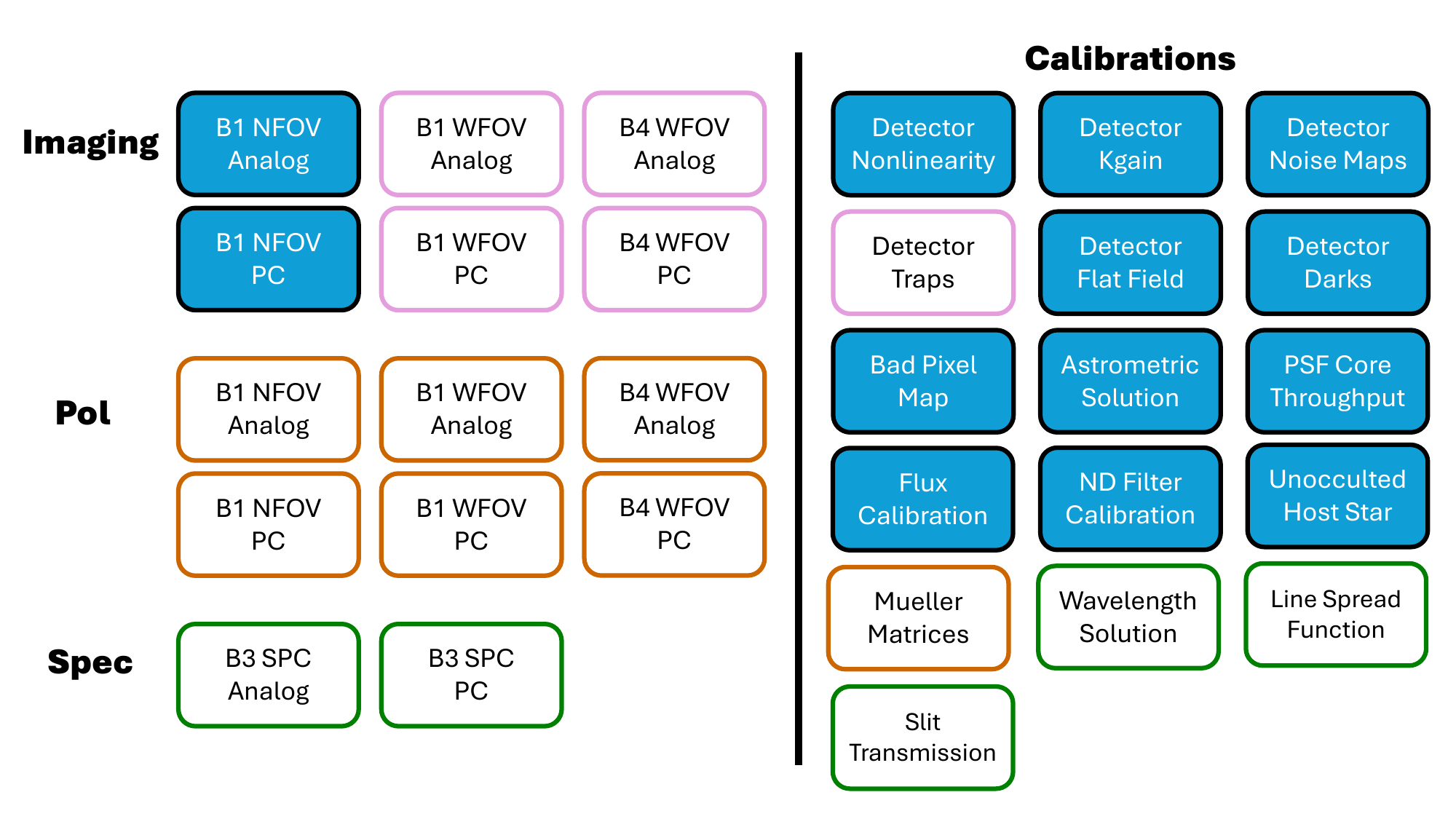}
    \caption{Different data products that are produced by \texttt{corgidrp} to support the first six months of planned Roman Coronagraph Observations. Boxes shaded in blue are needed to support the required mode of observation. B1, B3, and B4 refer to the filter bands 1, 3, and 4, respectively. NFOV and WFOV refer to the narrow and wide field of view modes (each with its own dedicated coronagraph). Analog and PC (Photon Counting) refer to the detector configuration. }
    \label{fig:data_products}
\end{figure}

With a preliminary plan for the first six months of observations established \cite{Wolff2026}, the initial data reduction needs for the Roman Coronagraph are also clear. We summarize all data products that will need to be produced by \texttt{corgidrp} to support the first six months of observations in Figure \ref{fig:data_products}. Below, we describe the various different data reduction needs and an overview of the data reduction process. We direct the reader to Ref.~\citenum{Bailey2023} for an overview of different modes, such as the wavelength coverage of the different bands, the field of view of the different modes, and the coronagraphs used.

\subsection{Band 1 Narrow Field Imaging with the Hybrid Lyot Coronagraph}
The most critical observing mode to support is narrow field imaging with the Hybrid Lyot coronagraph in Band 1. This mode will be used to satisfy the critical Threshold Technology Requirement to demonstrate the performance of the Coronagraph instrument. This mode will also be used to image exoplanets and exozodiacal dust in reflected light in the first six months of science observations \cite{Wolff2026}.

\texttt{corgidrp} will be needed to support these observations in three ways. First, the pipeline will be used to commission the instrument such as aligning the boresight of the Roman Coronagraph \cite{Chavez2026}. Second, calibration products (bad pixel maps, detector noise maps for generating synthetic darks, the Kgain value, and the detector flat field) produced by \texttt{corgidrp} will be used by the wavefront sensing and control algorithm to dig a dark hole on sky. Lastly, the science data taken by the Roman Coronagraph will be processed by \texttt{corgidrp}. The pipeline will measure the final sensitivity achieved after post-processing to determine whether the Roman Coronagraph has achieved its Threshold Technology Requirement. It will also search for and characterize any astrophysical sources imaged in visible light around the stars observed. 

The data reduction steps for this data are divided into stages which we call data levels. Ref.~\citenum{Millar-Blanchaer2024} provides an overview of each data level. L1 data is the raw data from the telescope, formatted into a 2-D image in a FITS file, and accompanied by header keywords. To produce a L2a frame from a single L1 data frame, \texttt{corgidrp} subtracts the bias value measured from bias regions that accompany each exposure, detects and marks pixels affected by cosmic rays, and corrects for nonlinearity in the pixel values. To produce L2b data products from L2a data products, the data are converted from detector counts to electrons using the commanded and measured detector gains, undergo dark subtraction, corrected for variations in the detector flat field, have bad pixels caused by cosmic rays and other detector artifacts masked, and optionally corrected for charge transfer inefficiency. For data taken in ``analog" mode where each frame has many photoelectrons, the L2a to L2b processing stage is done on each frame separately. Data taken in the photon counting mode will typically have either zero or one photon per pixel in the image, so they are binned together during this stage to produce photon-counted L2b images that contain multiple photons per image for ease of analysis in later steps. All data, including engineering data (e.g., dark hole digging frames), will be processed to the L2b stage.

L2b frames that are needed for stellar point spread function (PSF) subtraction are further processed. To process each L2b frame into an L3 frame, we discard any setup frames that are not needed for stellar PSF subtraction, divide each frame by its exposure time so that the pixel values represent electrons per second, add WCS information from the instrument's astrometric calibration, and crop the images from 1024x1024 to a much smaller footprint since most of the detector is not illuminated due to the field stop. 

L4 images require the combination of many images together to produce a combined frame that has the stellar PSF subtracted and stacks the exposures together. To do this, the input L3 dataset contains a subset of frames with ``satellite spots" produced intentionally by the deformable mirror (DM) to enable location of the star behind the coronagraphic mask for image registration. The satellite spot sequence is taken at the start of each coronagraphic sequence and consists of three parts with equal number of exposures: 1) images taken with no satellite spots for basic speckle subtraction, 2) images with satellite spots taken with a specific sinusoid placed on the DM, and 3) images with satellite spots with the negative version of the sinusoid from step two placed on the DM. The idea is that images from the two different satellite spot steps are co-added together to mitigate coherent interference between the satellite spots and the speckle field, and then the background frames are used to further suppress the speckle field to improve the efficacy of the star centering. The satellite spot images along with the regular coronagraphic images together are distortion corrected, have bad pixels filled in through interpolation, and are binned together to mitigate the effects of cosmic rays (exposures from the same visit with the same satellite spot configuration are binned together by default). The satellite spot frames are used to measure the star center in each visit (we assume that the star position is stable in a single visit due to the low order wavefront control system). The regular science images then have the stellar PSF subtracted using the KLIP algorithm \cite{Soummer2012} implemented in \texttt{pyKLIP} \cite{Wang2015} using a combination of reference star differential imaging (RDI) and angular differential imaging (ADI) \cite{Marois2006,Liu2004}. Additional stellar PSF subtractions are run with fake planets injected into the data to measure signal loss of off-axis sources in the field of view due to the PSF subtraction process. The images are then rotated so that North is aligned with the +y axis (``North-up") and all exposures are combined together. The final L4 image is a 3-D ``KL" cube, where the third axis in the number of principal component modes (often termed ``KL modes" in exoplanet imaging) used for stellar PSF subtraction. 

\begin{figure}
    \centering
    \includegraphics[page=6, width=\textwidth]{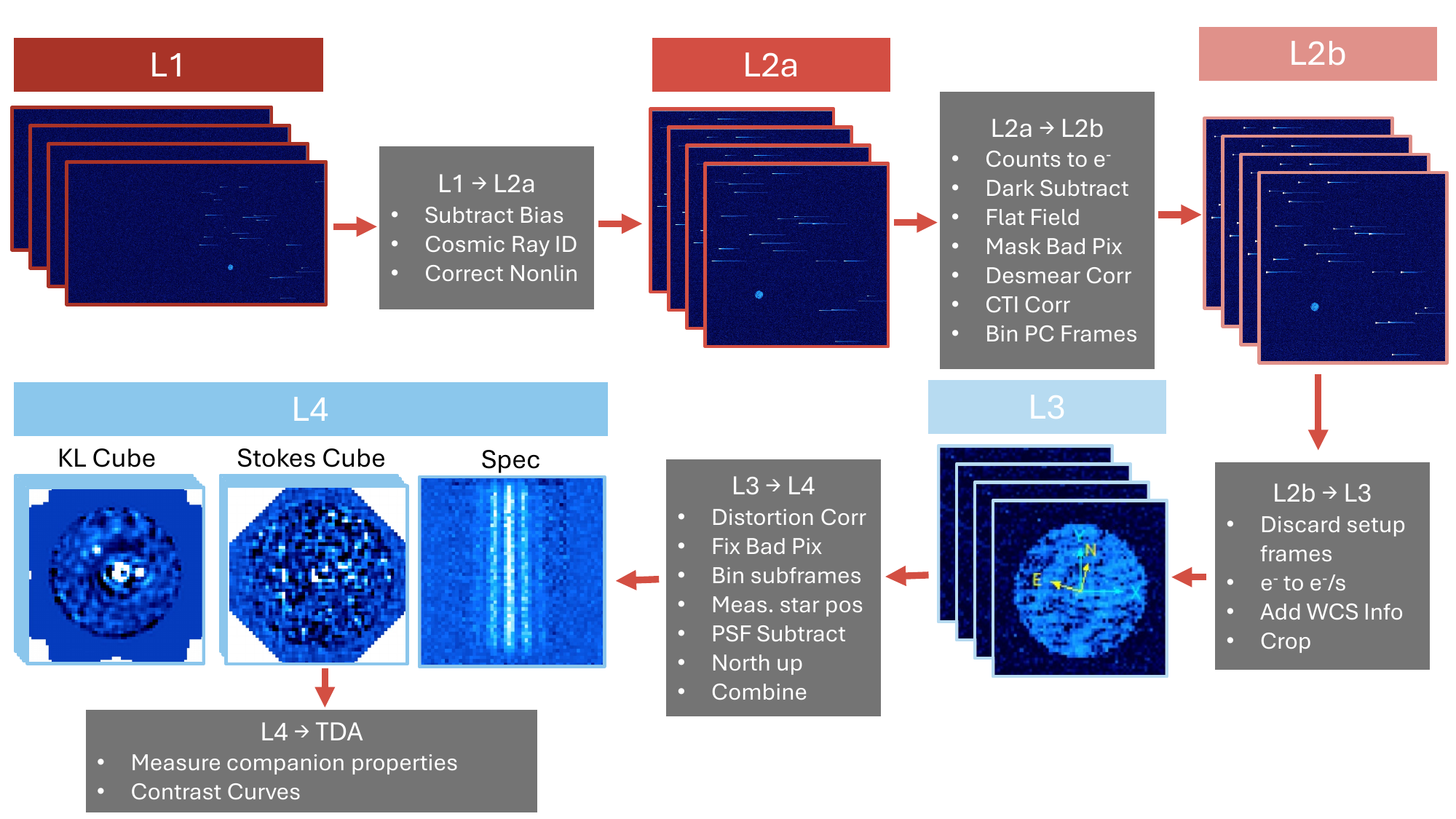}
    \caption{Data reduction steps for coronagraphic observations taken by the Roman Coronagraph. Simulated example data are shown in each official data level, which will be saved and publicly archived. }
    \label{fig:data_flow}
\end{figure}

The data flow is shown in Figure \ref{fig:data_flow}. \texttt{corgidrp} also provides functions to measure the sensitivity achieved in the L4 image (i.e., contrast curves) and the properties (i.e., astrometry and photometry) of any detected companions. These additional steps fall into the category of Tech Demo Analysis (TDA) and are not pipelined like the rest of the processing. However, \texttt{corgidrp} still provides functionality to make this analysis manually.

\subsection{Goal Modes}
Beyond the required Band 1 coronagraphic imaging in narrow field mode with the Hybrid Lyot coronagraph, the CPP has also planned several other modes of observations within the first six months of the science observation phase. These are considered ``goal" modes and have lower priority during commissioning. The goal modes are shown in Figure \ref{fig:data_products} without the blue shading. 

\textbf{Coronagraphic Imaging:} There are two other coronagraphic imaging modes that are goal modes: Band 1 and Band 4 imaging with the wide field of view shaped pupil coronagraphs. Like all coronagraphic data, the images can be taken in analog or photon counting mode. The data processing for these modes is nearly identical to the Band 1 narrow field of view imaging mode. The main differences are the usage of a different image crop size in L2b to L3 processing and different stellar PSF subtraction parameters. Additionally, there will be different calibrations needed (e.g., flat field, core throughput, ND filter, and absolute flux calibrations). 

\textbf{Polarimetry Mode:} In polarimetry imaging mode with the coronagraph, a Wollaston prism is inserted in the beam \cite{Groff2025}. The polarimetry mode and data reduction is discussed in detail in Refs.~\citenum{Anche2026} and \citenum{Groff2025}. For data processing, the main differences are that the images need to be divided by a polarized flat field in L2a to L2b processing, the images need to be cropped into a 3-D image cube with the two orthogonal polarization states from that frame's Wollaston prism in L2b to L3 processing, and the stellar PSF subtraction has two separate components: a polarimetric differential imaging routine is run in addition to the standard total intensity stellar PSF subtraction. The Stokes parameters are then recovered by demodulating the orthogonal polarization states through the system Mueller matrix. The resultant L4 data product is a data cube containing the Stokes parameters after stellar PSF subtraction, with the I component representing the total intensity stellar PSF subtraction and the Q and U components coming from the polarimetric differential imaging routine (Figure \ref{fig:data_flow}).

\textbf{Spectroscopy Mode:} In spectroscopy mode, a dispersive prism and slit are inserted in the optical beam to obtain slit spectra of a particular part of the field of view after the coronagraph \cite{Groff2025}. In the earlier data levels (before L3 to L4 processing), the main differences for spectroscopy mode are not dividing by a flat field and cropping down to a different size. The L3 to L4 data processing is significantly different in spectroscopy mode. No distortion correction is done. The satellite spot observation is taken with a narrow band filter so that the satellite spot is not smeared out by the prism dispersion. Additionally, only one satellite spot is seen through the slit, and the satellite spot position must be customized so that it is produced at the estimated position of the real companion of interest (i.e., on top of the companion). The single satellite spot position is measured both to determine the wavelength solution on the detector (since we know the central wavelength of the narrow band filter) given an input prism dispersion calibration, and to infer the star position through the known pattern commanded on the DM to produce the satellite spot. The image is stellar PSF subtracted through a simple RDI subtraction where the reference star speckles are scaled to match the spectrum of the target star. The image does not get rotated North up. Instead, each image is combined and a 1-D spectrum is extracted at the expected location of the companion. Note that both the 2-D image and the extracted 1-D spectrum are saved in the L4 data product (only the 2-D image is shown in Figure \ref{fig:data_flow}).

\subsection{Calibrations}
In order to process observations from L1 to L4 data products and to obtain measurements from the TDA analysis (e.g., contrast curves), a suite of instrument calibration measurements need to be made. A detailed calibration plan and justification of each calibration is described in Ref.~\citenum{Zellem2022}. The calibrations produced by \texttt{corgidrp} are shown in Figure~\ref{fig:data_products}. The blue shaded calibrations are needed for the required Band 1 narrow field of view imaging mode. Note that ``Detector Traps" is shaded as optional since the charge transfer inefficiency correction will be made on a best-effort basis. The detector calibrations (nonlinearity, Kgain, detector noise maps which produce synthetic detector darks, bad pixel maps) are generally agnostic of mode with the exception of flat field, which can depend on the observing band. The astrometric calibration (north angle, platescale, boresight offset, and field distortion) is also assumed to be independent of observing mode. The PSF core throughput measures how the off-axis PSF changes with location around the mask and depends on both the band and the coronagraphic mask used. The flux and ND filter calibrations depend on the wavelength band and whether the data is in spectroscopy mode or not (spectroscopy mode requires these values to be measured as a function of wavelength rather than a single value). The occulted observations of the target star are taken on a per star basis. Mueller matrices need to be computed for each combination of optics. Wavelength solutions and line spread functions are unique to each band. Slit transmission depends on both the band and the slit used.

All calibration products are also processed by \texttt{corgidrp}. Generally, a series of input L1 images will produce a single calibration file type, with a couple of exceptions. Their processing shares L1 to L2b processing steps with the coronagraphic observations whenever possible. Processing of calibration data beyond L2b generally differs from coronagraphic observations, and custom series of steps are used for calibration products that need additional processing beyond the L2b stage. For calibrations that are used during the L1 to L2b processing, they only experience a subset of the L1 to L2b processing steps along with custom processing to produce the calibration product.

\section{Pipeline Design Goals}
In this section, we discuss some high level pipeline design goals, both in terms of development and usage. 

\textbf{Built for Automation:}
To provide quicklook reductions as the data is being obtained, \texttt{corgidrp} should be able to be run by an automatic data processing system. The automatic data processing system, termed Data Integration Processor (DIP), should be able to call \texttt{corgidrp} with a set of data from a visit and \texttt{corgidrp} should figure out how to process the data, process it, and save it in a directory specified by the DIP system. Thus, \texttt{corgidrp} needs logic to determine how to process a dataset obtained from a single telescope visit and needs to be able to supplement the dataset with appropriate calibration files to perform the processing. 

\textbf{Flexibility in Usage:}
Based on our experience with other facilities, a data reduction pipeline should be flexible to handle the variety of different use cases beyond a preset automatic processing strategy. Some use cases include: 1) tweaking the data reduction of the automatic processing in small ways, 2) being able to reproduce the processing done by someone else, 3) calling specific steps of the data processing for a custom reduction, and 4) calling specific steps of the data processing by other software packages. We envision especially during commissioning and early science, the data reduction strategy will evolve quickly and \texttt{corgidrp} needs to be flexible enough to adapt. Additionally, community members beyond the CPP likely will want to process data in their own workflows. In summary, the individual data processing steps should have a clear API and can easily be used without a burdensome pipeline framework.

\textbf{Code Reuse:}
As seen in Figure~\ref{fig:data_products}, \texttt{corgidrp} needs to produce a lot of data products. As discussed in Section~\ref{sec:needs}, these data products share a lot of data processing. Like many other data reduction pipelines, we want to make sure we can maximize code reuse wherever possible to minimize development effort.

\textbf{Easy to Contribute:}
The Data Reduction and Simulations working group consists of a team of astronomers and engineers. Importantly, no one is dedicating their full time to pipeline development and there are very few software engineers. Thus, the pipeline needs to have a low barrier of entry for contribution. Currently, nearly 40 contributors have helped develop \texttt{corgidrp} and only 11 developers have made more than 50 commits. The majority of developers contribute a small number of features, but all these small contributions are a substantial part of the pipeline in the end. To minimize the barrier to contribution, we aimed to develop a simple core API and limit the amount of scaffolding code needed. We want to minimize the amount of time trying to learn a software framework and more time writing functions that do scientific analysis.

\textbf{Validated and Reproducible:}
As the official Roman Coronagraph data reduction pipeline, \texttt{corgidrp} is required to be extensively tested to demonstrate correctness and to validate pipeline behavior (e.g., pipeline API, data product formats). Additionally, we aim for all analysis from the Roman Coronagraph to be reproducible. As such, the pipeline needs an open-source testing suite that can be easily run by anyone. The testing suite needs to be extensive in order to test all the different kinds of analysis to be done (as shown in Figure \ref{fig:data_products}), and the testing suite needs to be thorough in order to validate both correctness and pipeline behavior.

\section{Pipeline Implementation}

\subsection{Pipeline Heritage}
The design and implementation of \texttt{corgidrp} was influenced by two pipelines that the lead developers of \texttt{corgidrp} had previously worked on or used. The first is the Gemini Planet Imager Data Reduction Pipeline (GPI DRP) \cite{Perrin2014, Perrin2016}. In particular, the GPI DRP was divided into a series of pipeline ``primitives" which are analog to our step functions, and it used text file recipes to record the data processing steps alongside a file-based calibration database that tracked and selected calibration files for each pipeline primitive. Furthermore, the GPI DRP data parser to automatically identify recipes given a set of input data products set the high level expectations of the walker module we discuss below. The second pipeline is the Keck Planet Imager and Characterizer Data Reduction Pipeline (KPIC DRP) \cite{Wang2021}, which implements in Python the idea of step functions, file-based calibration database, and the Image and Dataset classes to standardize data formats. \texttt{corgidrp} code for these three concepts is based on the implementation in the KPIC DRP.

\texttt{corgidrp} also depends on several open-source packages from the Python community. Throughout the code, \texttt{numpy} \cite{harris2020array}, \texttt{matplotlib} \cite{Hunter:2007}, \texttt{scipy} \cite{2020SciPy-NMeth}, and \texttt{Astropy} \cite{astropy:2013, astropy:2018, astropy:2022} are used. \texttt{photutils} \cite{bradley_2026_19636730} is used for several calibration routines. \texttt{statsmodels} \cite{seabold2010statsmodels} is used for nonlinearity calibration. \texttt{pyklip} \cite{Wang2015} is used for stellar PSF subtraction, rotation, centroiding, source finding, and extracting astrometry and photometry of off-axis sources. \texttt{emccd-detect} is used for mocking up fake data for testing.
The calibration database is based on \texttt{pandas} \cite{the_pandas_development_team_2026_21003741, mckinney-proc-scipy-2010}, which is also used for utility purposes across the code. The testing infrastructure uses \texttt{pytest} \cite{pytest}. \texttt{termcolor} is occasionally used for print messages in testing.

\subsection{Pipeline Components}
We discuss some key components of the pipeline design here. 

\begin{figure}
    \centering
    \includegraphics[page=7, width=0.8\textwidth]{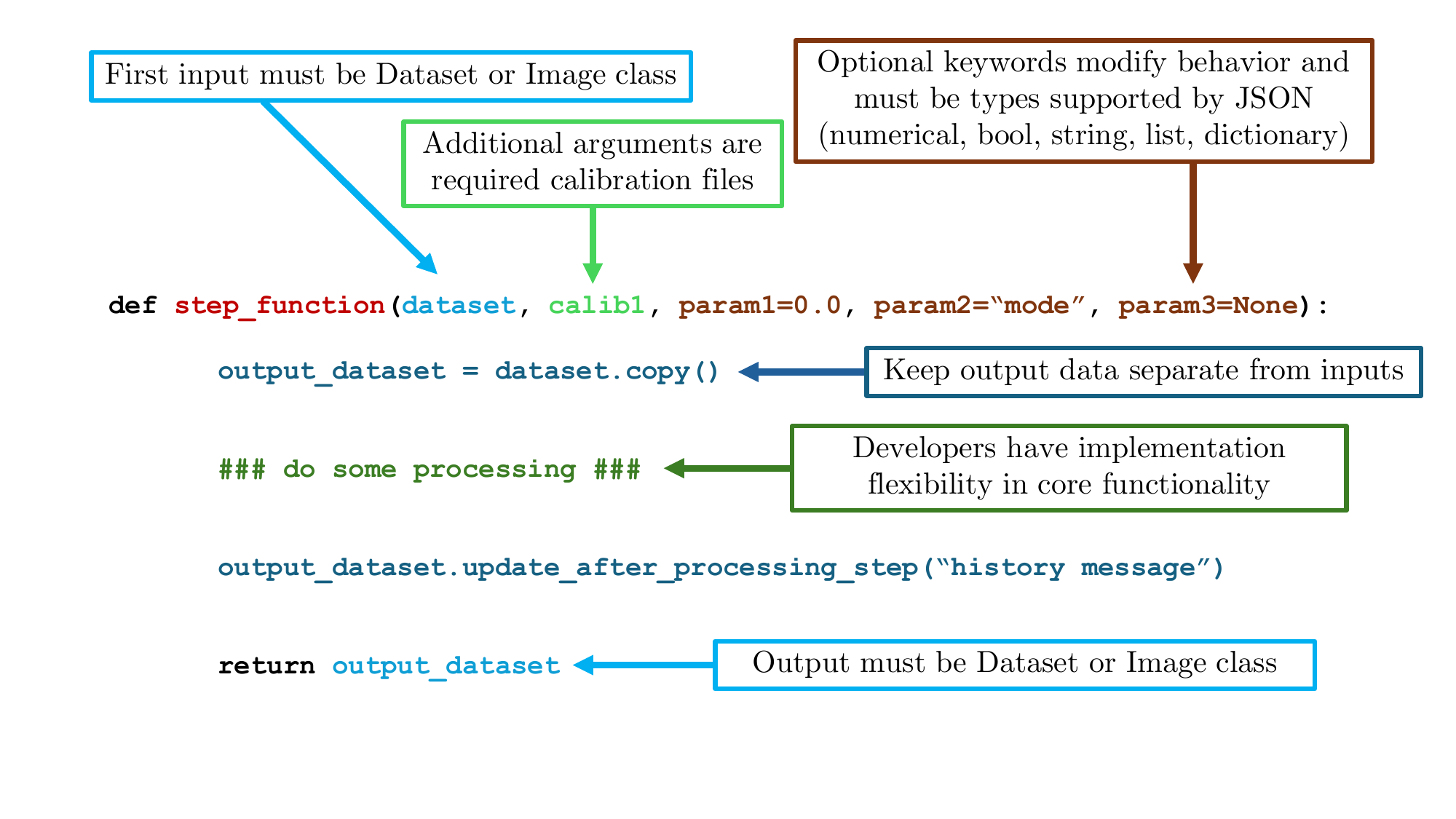}
    \caption{Components of a step function. Each step function takes an input \texttt{Dataset} or \texttt{Image} and returns an output \texttt{Dataset} or \texttt{Image}. The restrictions on function inputs allow step functions to be fully specified by JSON recipe files. The internals of the step function are extremely flexible to reduce the barrier for new developers.}
    \label{fig:stepfunc}
\end{figure}

\textbf{Data Classes:}
We use classes to standardize the data interface in the pipeline to ensure data is accessed, formatted, and saved in a common way. The \texttt{Image} class is the base class that represents any FITS file (raw or processed) from the Roman Coronagraph. The \texttt{Image} class standardizes the FITS data as such: a primary header data unit (HDU) that has a header but no data, the first extension HDU having the main data (e.g., the 2-D image) and a FITS header that \texttt{corgidrp} will write to, the second extension HDU with an error map, the third extension HDU containing a bitwise DQ map, and optional additional HDUs. The \texttt{Image} class also standardizes some header keywords, the save functionality, and how data are read in. A list of \texttt{Image} objects can be turned into an instance of the \texttt{Dataset} class, which preserves list functionality but enables batch updating for saving, updating headers, adding error terms, or splitting up the dataset by keywords. All normal coronagraphic data remain instances of the \texttt{Image} class the entire time through data processing, but each calibration data type is output as its own class, which is a subclass of the \texttt{Image} class. These subclasses enable the pipeline to easily differentiate between calibration data types, such as when picking calibrations to use.

\textbf{Step Functions:}
Step functions comprise the core functionality of the pipeline. The choice of using step functions enables \texttt{corgidrp} to follow a functional programming paradigm where the output of a step function only depends on the input (with almost no pipeline settings stashed away as attributes of a hypothetical pipeline class), making the code easier to test and validate. Figure \ref{fig:stepfunc} explains the anatomy of a step function. The inherent use of common functions and not classes for step functions enables flexible usage, including in the workflows of community members outside of the CPP (one just has to call the function). The function paradigm is something typical CPP members are familiar with, making it easy to add new pipeline functionality (one just has to write a function). The restricted API of step functions (ordering and types of inputs allowed, only one kind of output allowed) enables step functions to be chained together for easy pipeline processing (the output of one function is the input of the next). When combined with the fact that all Roman Coronagraph data can be loaded in as instances of the \texttt{Image} class with common read/write functions, the framework to pipeline process data only needs to know to use the features of the \texttt{Image} class to pass data in and out of step functions and to load/save data, reducing complexity of the automation framework. 

\begin{figure}
    \centering
    \includegraphics[page=8, width=0.8\textwidth]{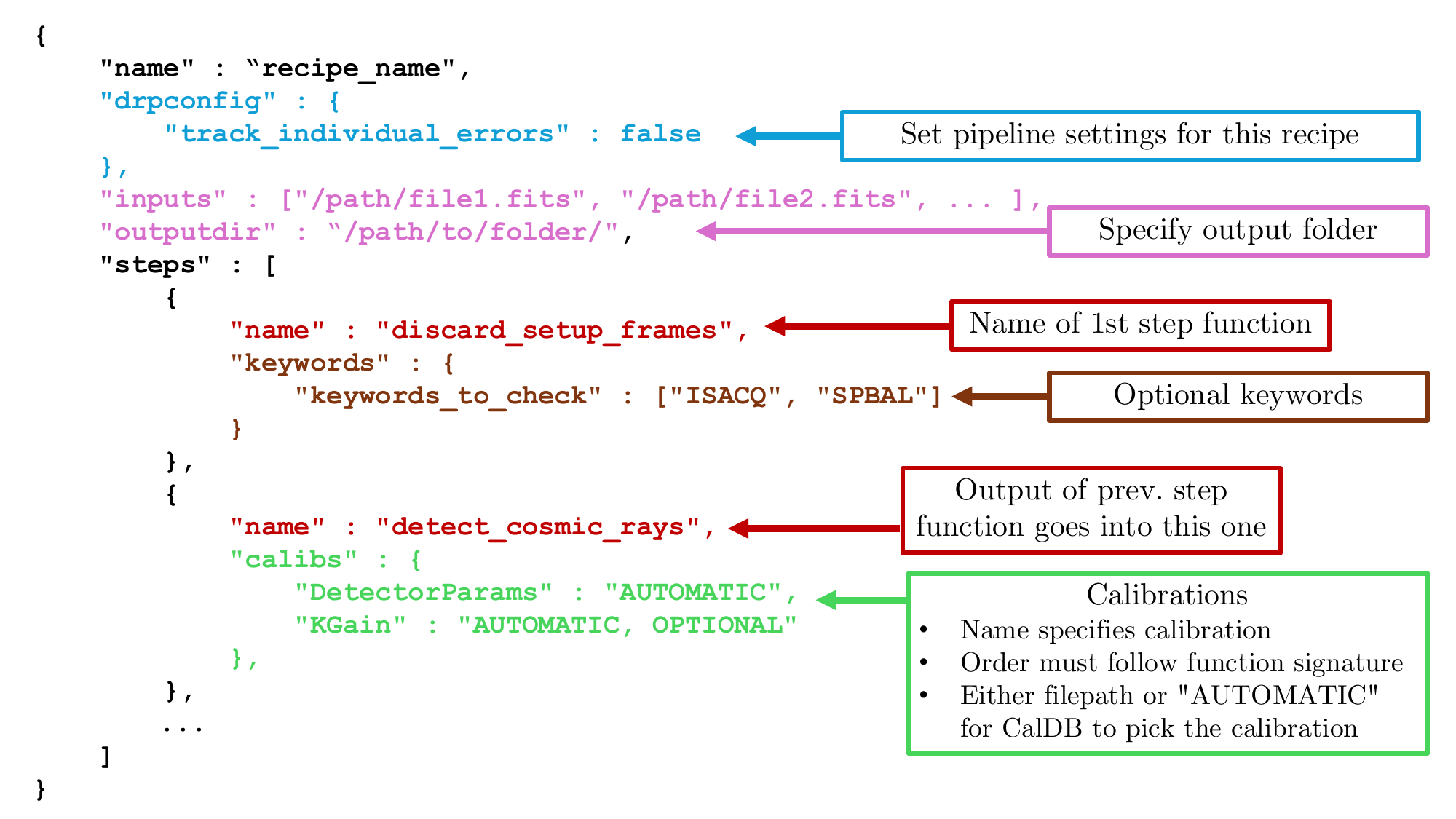}
    \caption{Components of a recipe. This diagram highlights how a JSON recipe is able to modify pipeline settings, determine the input data and output directory, specify the step functions to run, specify any tweaks to step function parameters, and specify calibration data needed.  }
    \label{fig:recipe}
\end{figure}

\textbf{Recipes:}
By using a functional programming paradigm for the step functions, our data processing recipes are also naturally straightforward. Figure \ref{fig:recipe} explains the structure of our JSON-based recipes. Given the minimal state that is hidden in pipeline attributes, a single human-readable JSON recipe details the full data processing of a pipeline product. This improves reproducibility as these recipes are saved into the FITS file headers, so a user can reload the exact recipe from the header and use it to reproduce the data processing. Users can also write their own recipes from scratch or modify a recipe to tweak knobs of individual step functions. The supported datatypes of JSON files restrict the datatypes of step function keywords. Because the required arguments to functions are either instances of the \texttt{Dataset} or \texttt{Image} class and because these classes have standardized functions to load data from disk, filepath strings and strings to denote the calibration subclass are sufficient in the JSON recipes for the pipeline to know how to load and pass these data into the step functions. \texttt{corgidrp} comes with a folder of template recipes that represent all of the supported data processing and other common data processing procedures. The templates are merely missing a list of input data files, the output data directory, and the file paths to required calibration files needed for various step functions. 

\begin{figure}
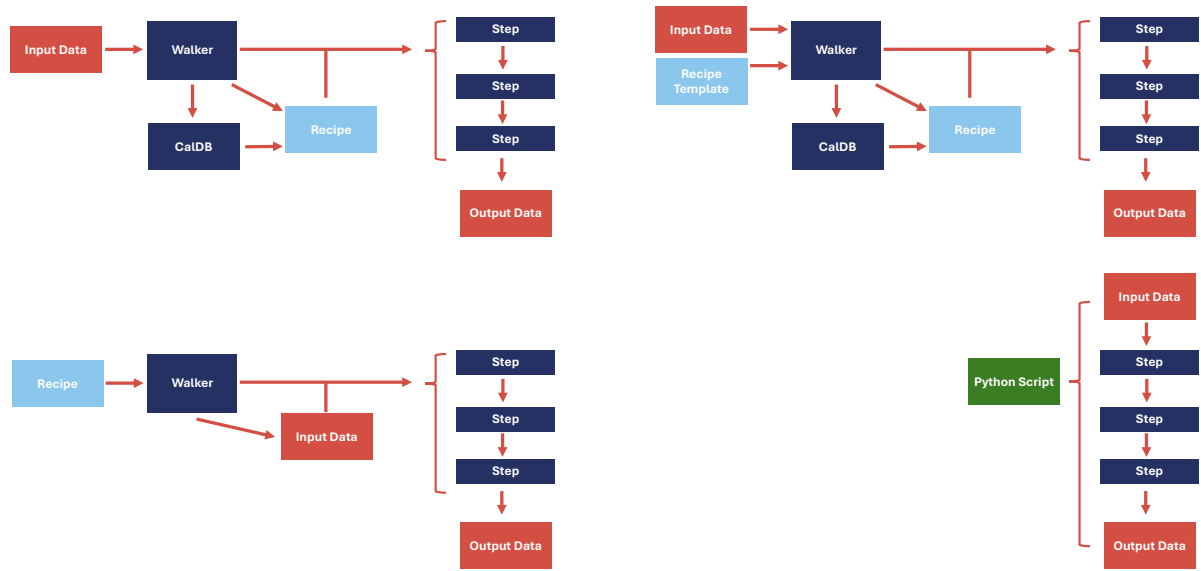

    \centering
    \includegraphics[page=9, width=0.49\textwidth, trim={1cm 0cm 2cm 3cm}]{DRPDiagrams.pdf}
    \includegraphics[page=10, width=0.49\textwidth, trim={1cm 0cm 2cm 3cm}]{DRPDiagrams.pdf}
    \includegraphics[page=11, width=0.49\textwidth, trim={1cm 0cm 2cm 3cm}]{DRPDiagrams.pdf}
    \includegraphics[page=12, width=0.49\textwidth, trim={1cm 0cm 2cm 3cm}]{DRPDiagrams.pdf}
    \caption{Four different ways we envisioned data can be processed with \texttt{corgidrp}. Top left: input data is fed into the walker and the walker sets the recipe and processes the data. Top right: Input data with a recipe template is fed into the walker and the walker processes the data after populating the recipe template. Bottom left: A recipe that fully specifies everything is given to the walker and the walker loads that data and processes it according to the recipe. Bottom right: a user script calls the step functions and reduces the data without any pipeline framework. }
    \label{fig:pipeline-options}
\end{figure}

\textbf{Walker and Automation API:}
The \texttt{walker} module in \texttt{corgidrp} is the pipelining software layer. As of writing, this module is less than 1000 lines long, and is part of our philosophy of minimizing the amount of pipeline framework a developer needs to learn to contribute to the code. Inside the walker is the code to guess which recipe template to use given an input list of data, populate a recipe template into a full recipe, and run the recipe. Figure \ref{fig:pipeline-options} shows some options for how we envision data can be processed with the walker and the step functions. The top two scenarios are the main scenarios that the CPP will use with the DIP. The DIP calls \texttt{corgidrp} and the \texttt{walker} module through the \texttt{ops} interface. Only the DIP is the intended user of the \texttt{ops} module.

\textbf{Calibration Database:}
We put all the responsibility of calibration file management in the \texttt{caldb} module, which implements a file-based database of calibration files that lives on a machine. The calibration database is one of the few components of the pipeline that does preserve state, and thus does not follow the functional programming paradigm. This is necessary because it keeps track of all calibration files produced on the machine so that it can be used for further processing. Inspired by the GPI DRP, it is also able to scan a folder and load all valid calibration files into the database, so that a user can download a package of calibrations onto their machine and load them into \texttt{corgidrp} easily. The \texttt{caldb} has a series of rules it follows that when given input data to process and a requested calibration type, it finds the ``best" calibration file available. This functionality simplifies design on two ends. For the walker and recipe, it can rely on specifying that calibration files will be automatically identified at run time by the calibration database. For step function development, a developer does not need to think about how to query the calibration database and selection rules, and instead merely needs to specify required calibrations as inputs to the step function.

\subsection{Development Coordination}
\texttt{corgidrp} is tracked by git and hosted on Github at \url{https://github.com/roman-corgi/corgidrp}. The ``develop" and ``main" branches are both protected and cannot be directly pushed to. Following the Gitflow workflow, new releases generally are done on the ``main" branch, while ``develop" hosts new changes for the next release. Project release goals are tracked on the Github issue tracker, and we encourage all users to make issues of any issues they encounter on the issue tracker. Contribution instructions exist in the README. Contributors are instructed to make a feature branch to develop their code and then submit a pull request to the ``develop" branch when ready. All pull requests are reviewed line-by-line by the main maintainers of the codebase. Automated unit tests are also run on every pull request, and the reviewers can choose to run some or all of the end-to-end tests manually to verify certain features (see next section for a discussion of testing). The CPP currently runs biweekly Data Reduction and Simulations telecons where code development is discussed and coordinated as well.  

\subsection{Testing Suite}
The testing philosophy for \texttt{corgidrp} is driven by the desire to test every data reduction scenario we expect to encounter for real data: we want to test that we can produce all of the data products shown in Figure \ref{fig:data_products} given raw L1 data. This is one of the main goals of what we term ``end-to-end" tests. They test a full use case (e.g., those shown in Figure \ref{fig:pipeline-options}). This stems from our testing philosophy that the large majority of bugs in software come from the integration of software components where a change in one component can propagate unintended consequences in another component. End-to-end tests that exercise the data reduction pipeline in all its various modes will be effective at identifying these bugs. These end-to-end tests require several GBs of simulated data and take over an hour to run, so they are not practical for efficient pipeline validation. Thus, these are run as-needed when reviewing new pull requests and before the release of a new pipeline version.

For faster tests that can run automatically on every pull request, we also have an extensive suite of unit tests. These unit tests only test the functionality of individual functions and modules rather than how they work together on realistic data. Developers are all generally expected to write at least one unit test for every new component of code, and reviewers can request additional unit tests when needed. As of writing, \texttt{corgidrp} has 391 unit tests, and running them takes roughly 35 minutes. These tests have been effective enough at catching errors so that the extensive end-to-end tests have not needed to be integrated into automated pull request testing.

The documentation and verification of software behavior is a second key goal of our testing philosophy. This is important as \texttt{corgidrp} needs to be run automatically in the DIP and the data products produced need to be archived and made publicly accessible. This means that certain code behaviors must remain as expected and the output data products (all of them in Figure \ref{fig:data_products}) need to follow given specifications. The unit tests help with this to some degree by testing key interfaces such as the \texttt{ops} module used by the DIP. The end-to-end tests also play an important role in validation of the data products produced by the pipeline. All data format specifications are written into the documentation pages of \texttt{corgidrp}, which are also available on readthedocs\footnote{\url{https://corgidrp.readthedocs.io/en/main/data_formats/index.html}}. After producing a suite of data products in the end-to-end testing, a final end-to-end test automatically generates data format pages for each data product type. Each of the data format pages is compared to the static versions of the same pages and any non-trivial differences will cause the test to fail. To make the test pass, the user can fix the test or update the docs pages, which can be source-controlled on git, so that there are no longer any non-trivial differences. Thus, all changes in data format are also naturally tracked by git, ensuring a history of documented data format changes.

\section{Future Development}
As of writing, version 5.0 of \texttt{corgidrp} has been released. This is intended to be the version of the pipeline used for initial commissioning and science. We envision there will likely be minor bug fixes in the coming months, resulting in minor updates to version 5.0. One of the major goals before launch of the Roman Space Telescope is further end-to-end development. Many of our end-to-end tests can be upgraded with more realistic data sequences, given that \texttt{corgisim} has improved \cite{Zhang2026} and given that observing sequences for commissioning and the first six months are well defined \cite{Wolff2026}. Furthermore, some data products lack full testing starting at raw L1 data, and end-to-end tests still need to be written for them. While this work will not add pipeline functionality, we hope it will exercise most of the bugs in the pipeline before real data starts coming in.

We anticipate that once we start getting real data, data processing algorithms likely will need to be improved. We anticipate a future version 6.0 release in the next calendar year to address some of the unanticipated challenges of real data. As part of this, we hope that community members beyond the CPP will use our pipeline and contribute upgrades as well. 

To improve community engagement, another aspect of the pipeline development that needs more work is the documentation site. Currently, the site does a detailed job of documenting data types. However, the documentation lacks tutorials and instructions for how to use the data products to analyze data. A concerted effort on this front should improve the accessibility and utility of the pipeline. 

\acknowledgments 
The authors wish to acknowledge Sergi Hildebrandt and Sarah Betti for contributions to \texttt{corgidrp}.
This material is based upon work supported by NASA under award Nos. 80NSSC24K0087, 80NSSC25K0373, and 80NSSC25K0364. 
Alexis Lau and Sophie Noiret acknowledge the support by the European Union (ERC, ESCAPE, project No. 101044152). Views and opinions expressed are, however, those of the author(s) only and do not necessarily reflect those of the European Union or the European Research Council Executive Agency.
The research was carried out in part at the Jet Propulsion Laboratory, California Institute of Technology, under a contract with the National Aeronautics and Space Administration (80NM0018D0004).
\bibliography{report} 
\bibliographystyle{spiebib} 

\end{document}